\documentclass[aps,pra,reprint,superscriptaddress,amsmath,amssymb,floatfix]{revtex4-2}

\usepackage{graphicx}
\usepackage{dcolumn}
\usepackage{bm}
\usepackage{braket}
\usepackage{hyperref}
\usepackage{xcolor}
\usepackage{etoolbox}
\usepackage{wrapfig}
\usepackage{svg}
\usepackage{pgf}

\usepackage[font={sf, small}, justification=RaggedRight,labelfont=bf]{caption}

\usepackage[cm]{sfmath}

\begin{document}

\title{Alignment-free photonic interconnects}

\author{Saumil Bandyopadhyay}
\affiliation{Research Laboratory of Electronics, MIT, Cambridge, MA 02139, USA}

\author{Dirk Englund}
\affiliation{Research Laboratory of Electronics, MIT, Cambridge, MA 02139, USA}

\begin{abstract}
\noindent Next generation optoelectronic systems will require the efficient transfer of optical signals between many  discrete photonic components integrated onto a single substrate. While modern assembly processes can easily integrate thousands of electrical components onto a single board, photonic assembly is far more challenging due to the wavelength-scale alignment tolerances required. Here we address this problem by introducing an alignment-free photonic coupler robust to $x,y,z$ displacement and angular misalignment. This alignment-free coupler engineers a translationally-invariant evanescent interaction between  waveguides by intersecting them at an angle, which enables a lateral and angular alignment tolerance fundamentally larger than non-evanescent approaches such as edge coupling. We show that our approach can function as a universal photonic connector interfacing photonic integrated circuits and microchiplets across different platforms. As a potential use case, we describe a new type of self-aligning photonic circuit board enabled by our approach that greatly simplifies assembly of complex optoelectronic systems.
\end{abstract}

\maketitle

Interconnects are indispensable in electronics, enabling circuit components with diverse functionalities to be assembled together into complex systems. The same need for system integration exists in optics, but progress is held back by a lack of reliable and easy-to-align photonic interconnects between components at the inter-chip and intra-chip levels. 

The desired attributes in optical interconnects are single-mode propagation, low loss, ease of manufacturing, and compatibility with high-density electrical interconnect technologies such as flip-chip bonding. Tapered adiabatic couplers  \cite{dangel_polymer_2018, soganci_flip-chip_2013, tiecke_efficient_2015} are a common choice for interfacing to waveguides on a photonic integrated circuit (PIC), but at the expense of high demands on nanofabrication, wavelength-scale and mostly manual alignment, and difficult scaling. Alternative approaches have been proposed, including photonic ``wirebonds'' that connect PICs through flexible polymer waveguides \cite{lindenmann_connecting_2015, lindenmann_photonic_2012, dietrich_situ_2018}, integrated optical microlenses \cite{scarcella_pluggable_2017, mangal_ball_2020}, pitch-reducing interposers and fiber arrays \cite{hwang_128_2017, pashkova_development_2019}, and bulk optical components such as parabolic reflectors microfabricated into polymer films \cite{ogunsola_chip-level_2006, lin_low-cost_2013, yu_optical_2020}. Small numbers of PICs can be also be connected by conventional fiber with edge coupling \cite{Almeida:03} or grating coupling \cite{Waldhausl:97, taillaert_out--plane_2002}, but scaling to tens of channels becomes extremely challenging.

Compared to electrical contacts, all of these photonic approaches require alignment that is orders of magnitude more demanding in angle and displacement. This disparity in required tolerances has presented a critical roadblock to scaling photonic systems. 

We propose to solve these problems by introducing a photonic interconnect technology that is largely insensitive to misalignment. Our approach relies on the interaction between two waveguides crossing at an angle, which is optimized for efficient evanescent coupling  at their intersection. This coupler is \textit{invariant} to translational misalignment, as the intersection between two lines is invariant to any in-plane translation $\Delta r_{||}$. In addition to translational invariance, the coupling efficiency is far more insensitive to angular misalignment $\Delta \theta$ than conventional approaches such as edge coupling. From detailed analytical models with closed-form expressions for coupling efficiency as a function of device materials and geometries, we present efficient design methods and show alignment tolerance that is fundamentally impossible to achieve with edge or grating couplers, which exhibit a fundamental tradeoff $(\Delta r_{||} \Delta \theta)_{3~\text{dB}}=\lambda/\pi n$, where $\lambda/n$ is the wavelength in an effective index $n$. Finally, we extend our scheme to also allow out-of-plane tolerance $\Delta z$ through a cantilevered alignment-free coupler design.

\begin{figure*}[t]
    \centering
    \includegraphics[width=\textwidth]{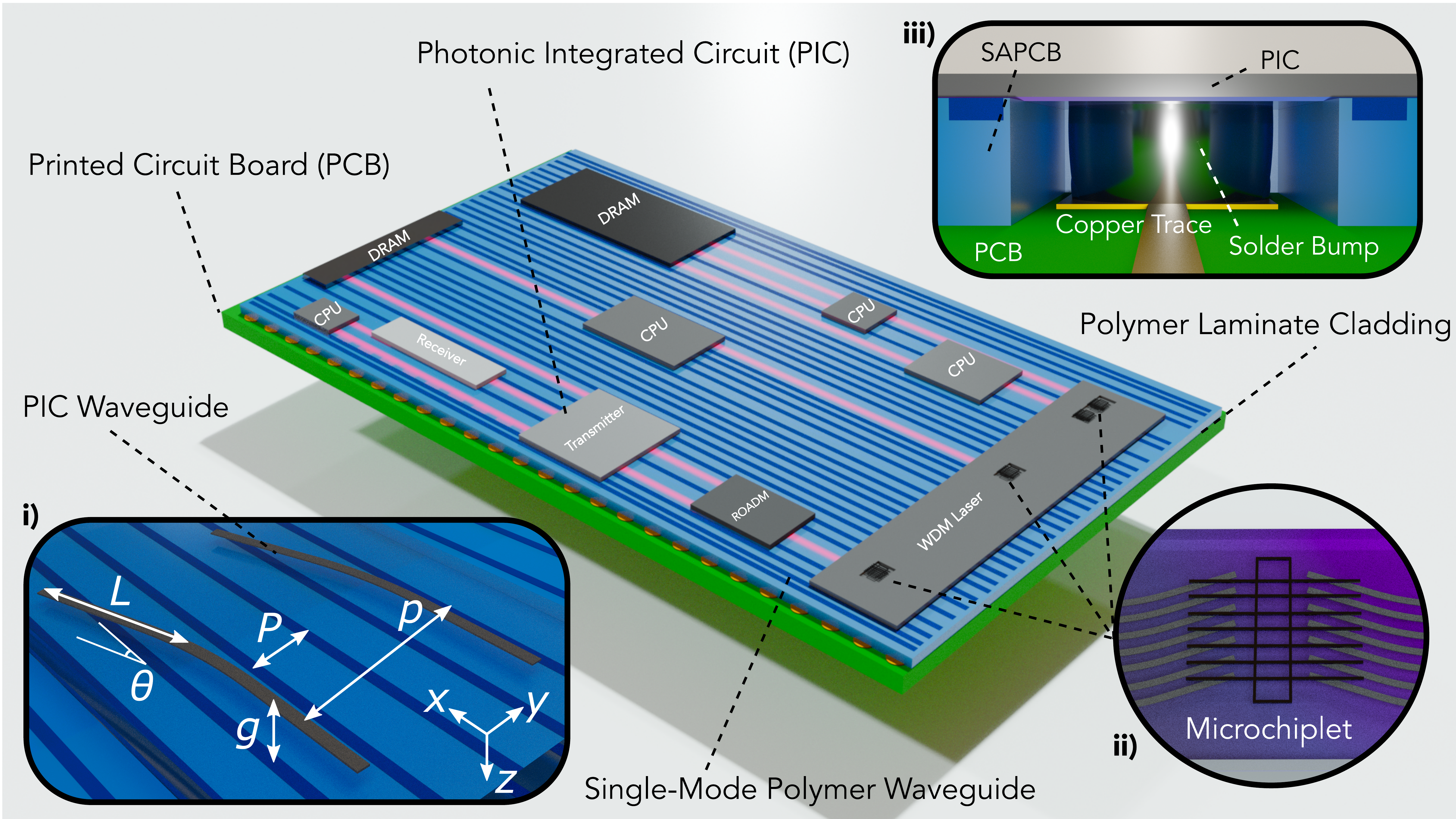}
    \caption{The SAPCB consists of a polymer-laminate film bonded onto an electrical PCB. PICs are flip-chip bonded to the polymer film, which includes a linear, closely-spaced array of single-mode waveguides that carry signals between chips. The polymer waveguides can be spaced with a pitch $P$ that is  smaller than the waveguide pitch $p$ on the PIC, which permits arbitrary placement of PICs and enables the transmit port of a chip to be paired easily to the receiver port on another. i): The SAPCB consists of efficient, board-level optical interconnects by making use of an alignment-free ``hockey stick'' coupler that intersects the polymer waveguides at an angle $\theta$. This approach makes the efficiency of our architecture insensitive to in-plane displacements and permits coupling over a wide range of waveguide pitches. Additionally, intersecting the two waveguides at an angle eliminates the requirement to place PICs onto the SAPCB with sub-micron placement accuracy. ii) The alignment-free coupler also simplifies  ``pick-and-place'' integration of microchiplets into PICs, which enables the introduction of gain, detectors, and single-photon sources into a single chip. iii): Electrical connections can be made in our architecture by punching holes through the polymer film, which permits bump bonding to pads on the electrical PCB.}
    \label{concept}
\end{figure*}

As a particular use case enabled by our approach, we introduce a ``self-aligning photonic circuit board'' (SAPCB)  that unifies photonic integrated circuits, microchiplets, and electronics onto a single optoelectronic substrate. Similar to other optical PCBs \cite{dangel_polymer_2018, soganci_flip-chip_2013, mangal_ball_2020, lin_low-cost_2013, yu_optical_2020}, the SAPCB's waveguides are made of polymer, making them easy and scalable to fabricate. However, unlike other optical PCBs, which typically require defining complex waveguide routing in polymer to carry signals between components, the SAPCB consists solely of a linear array of waveguides, making it far easier to manufacture. By fabricating an array of waveguides with variable widths, one can create a universal connector to match PIC waveguides of varying materials or dimensions, thereby facilitating the assembly of diverse photonic and electronic components into high-density systems. 

Figure \ref{concept} illustrates potential applications of the alignment-free coupler. We envision an SAPCB comprising a polymer-laminate film bonded to an electrical PCB onto which the photonic chips are placed. The polymer film incorporates a closely-spaced, linear array of single-mode waveguides used for optical interconnections between PICs. Waveguides on the polymer film are evanescently coupled  to those on the PIC over a vertical gap size $g$. 

The critical requirement of the SAPCB architecture is a board-to-PIC coupler with high efficiency and alignment tolerance. Approaches such as adiabatic coupling or edge coupling have demanding requirements ($<5~\mu$m) for alignment precision \cite{dangel_polymer_2018, soganci_flip-chip_2013, tiecke_efficient_2015}; moreover, the strict alignment tolerance requires that the photonic circuit and  substrate be co-designed to ensure the placement of the polymer waveguides are matched to those of the PICs with micron-scale precision. To solve these problems, we introduce the alignment-free, ``hockey stick'' coupler illustrated in Fig \ref{concept}(i). These devices consist of a PIC waveguide that runs parallel to the polymer film until taking a turn to intersect the polymer waveguides at an angle $\theta$ and length $L$. The angle $\theta$ is chosen to efficiently transfer optical power through the interaction of their evanescent fields. Off-axis alignment is usually considered undesirable for optical coupling; here, our approach intentionally applies it to achieve one critical benefit: this geometry is \textit{invariant} to any longitudinal displacement $\Delta x$ and any transverse displacement $\Delta y < L \sin \theta$. Moreover, the transverse displacement tolerance can be increased arbitrarily by increasing the length of the coupler $L$. 

Angled coupling introduces other benefits during assembly. Suppose the polymer and PIC waveguides have differing pitches $P, p$, respectively.
No matter their respective pitches, as long as the two waveguides are coarsely aligned within $L \sin \theta$, they will always intersect at some point with no transmission penalty. The SAPCB could therefore serve as an off-the-shelf, universal connector interfacing PICs of different designs and with differing port locations. The only restriction on the polymer waveguide pitch is that $P$ must be smaller than $L \sin \theta$, which ensures that no waveguide on the PIC couples to more than one polymer waveguide.

In addition to board-level assembly, the alignment-free coupler also enables simplified ``pick-and-place'' integration of microchiplets into photonic circuits. Microchiplets, which are miniaturized photonic chips integrated into larger circuits, have recently drawn interest as an approach for integrating gain \cite{theurer_flip-chip_2019, op_de_beeck_heterogeneous_2020, zhang_transfer-printing-based_2018, de_groote_transfer-printing-based_2016}, photodetectors \cite{piels_low-loss_2014, shen_hybrid_2017}, or single-photon sources \cite{wan_large-scale_2020} into PICs. This integration is illustrated in Figure \ref{concept}(ii), where the PIC backbone has windows etched into the cladding for coupling chiplets to the circuit. Finally, the SAPCB is also compatible with state-of-the-art electrical interconnect technologies such as flip-chip bonding. We illustrate this compatibility in Figure \ref{concept}(iii), which shows how holes can be punched in the polymer film to enable bump bonding between the PIC and PCB.

\begin{figure}
    \centering
    \includegraphics[width=3.4in]{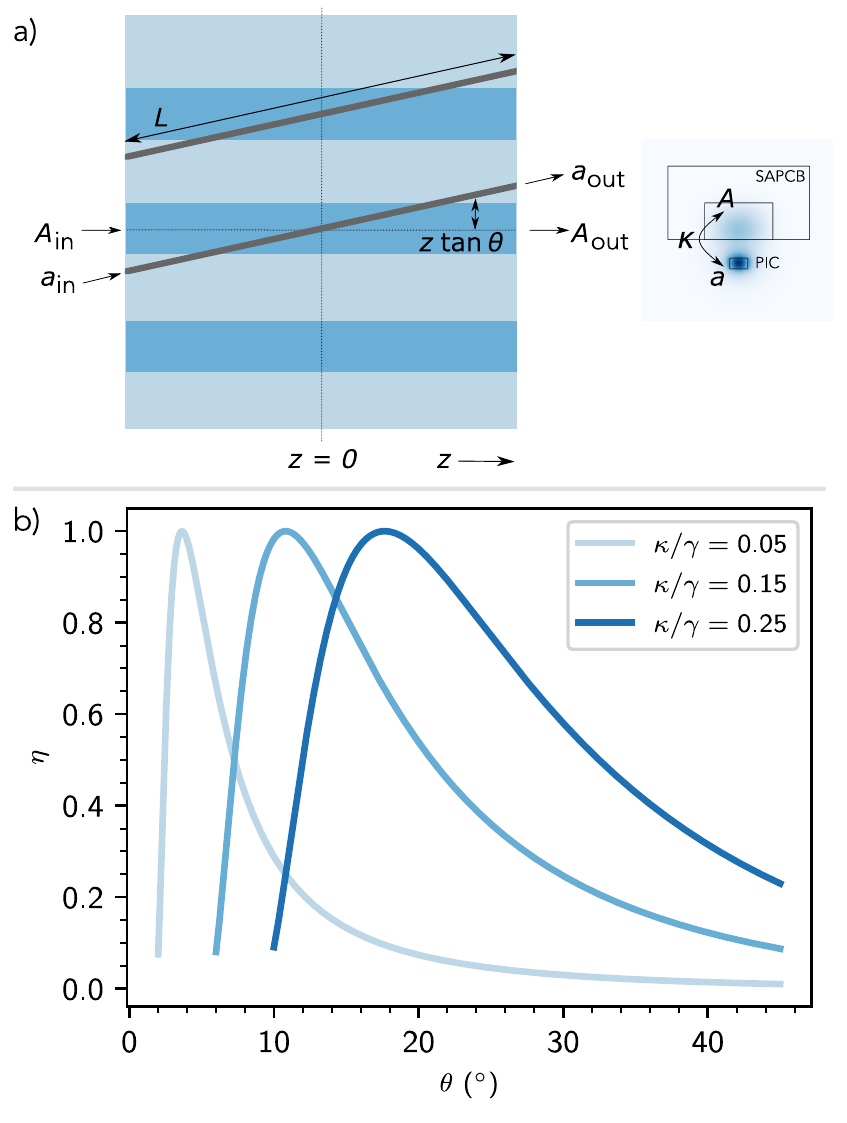}
    \caption{a) The alignment-free coupler can be modeled as two waveguides weakly coupled vertically by an evanescent interaction strength $\kappa$  to one another at an off-axis angle $\theta$. At an arbitrary point $z$ along the propagation, the coupling constant $\kappa$ will decay exponentially by the vertical offset $z \tan \theta$ with a characteristic decay length $\gamma$, i.e. $\kappa(z) = \kappa_0 e^{-\gamma |z| \tan \theta}$. b) The theoretical power transfer efficiency $\eta$ of the alignment-free coupler vs. $\theta$ for varying values of $\kappa/\gamma$. At small values of $\theta$, $\eta$ will oscillate rapidly from minimum to maximum power transfer. The alignment-free coupler should not be used in this regime and it is omitted from the plot for clarity.}
    \label{cmt}
\end{figure} 

\section*{Results}

\textbf{\textit{Theory}}: We begin by analyzing the dynamics of the alignment-free coupler using a coupled mode theory approach \cite{sun_adiabaticity_2009, marom_relation_1984} (Fig.~\ref{cmt}a). Consider two waveguides weakly coupled vertically to one another at an off-axis angle $\theta$ and intersecting at $z = 0$. When the two waveguides intersect one another, their interaction can be described by a coupling constant per unit length $\kappa$ and a wavevector  mismatch $\Delta k$. At an arbitrary $z$, $\Delta k$ remains unchanged, but $\kappa$ exponentially decays with the transverse offset $|z| \tan \theta$ \cite{chrostowski_hochberg_2015}. The waveguide coupling can therefore be modeled as $\kappa(z) = \kappa e^{-\gamma |z| \tan \theta} = \kappa e^{-\gamma^\prime |z|}$, where $\gamma^\prime=\gamma \tan \theta$ describes the decay of $\kappa$ with transverse offset per unit length.

The following coupled mode equations describes the interaction between the two waveguides with amplitudes $a$ (PIC) and $A$ (SAPCB):
\begin{align}
    i \frac{\mathrm{d}}{\mathrm{d}z}
    \begin{bmatrix}
        a_\text{out} \\
        A_\text{out}
    \end{bmatrix} &=
    \begin{bmatrix}
         \frac{2 \pi}{\lambda} n_\text{PIC} &  \kappa e^{-\gamma^\prime |z|} \\
         \kappa e^{-\gamma^\prime |z|} &  \frac{2 \pi }{\lambda} n_\text{SAPCB}
    \end{bmatrix}
    \begin{bmatrix}
        a_\text{in} \\
        A_\text{in}
    \end{bmatrix} 
\end{align}
The dimensions of the two waveguides are chosen to ensure the effective mode indices $n_\text{PIC}, n_\text{SAPCB}$ are equal, i.e. $\Delta k = 2 \pi(n_\text{SAPCB} - n_\text{PIC}) / \lambda = 0$. Analytically solving these equations provides the power transfer efficiency $\eta$ at an angle $\theta$ when the effective indices are matched:
\begin{equation}
    \label{theory}
    \eta = \left \lvert\frac{A_\text{out}}{a_\text{in}}\right \rvert^2 = \sin^2 \left ( \frac{2 \kappa}{\gamma \tan \theta} \right )
\end{equation}
$\eta$  reaches unity when the argument of the sine function is $\pi/2$, i.e. $\theta_\text{opt} = \arctan [(4/\pi) \kappa / \gamma]$. The 3-dB angular tolerance $\Delta \theta$ is therefore:
\begin{align}
    \Delta \theta = &\arctan \left [\frac{8 \kappa}{\pi \gamma} \right ] - \arctan \left [\frac{8 \kappa}{3\pi \gamma} \right ] \\ &\approx \left ( \frac{16}{3 \pi} \right ) \frac{\kappa}{\gamma} = \frac{4}{3} \theta_\text{opt} 
\end{align}
where we make the approximation $\arctan \theta \approx \theta$ for small coupling angles $\theta$.

Figure \ref{cmt}b plots the theoretical transmission efficiency $\eta$ vs. angle $\theta$ for varying values of $\kappa/\gamma$. In addition to potentially arbitrary lateral tolerance, depending on the value of $L$, the alignment-free coupler has high angular tolerance. This coupling scheme therefore has two major advantages over conventional optical couplers:
\begin{itemize}
    \item \textbf{High angular tolerance}: The $1/\tan \theta$ dependence of $\eta$ produces a large angular tolerance $\Delta \theta = (4/3) \theta_\text{opt}$. Moreover, $\eta$ has a long tail that ensures modest coupling even at very large angular errors, greatly simplifying initial alignment. Coupling the waveguides more strongly (increasing $\kappa/\gamma$) further increases $\Delta \theta$. 
    \item \textbf{Robust design}: no matter the values of $\kappa, \gamma$, the coupling efficiency reaches unity at some angle. Fabrication-induced variation in $\kappa$ can therefore \textit{always} be corrected during alignment. No matter the design, the angled coupler allows efficient power transfer by rotating one waveguide relative to the other. By contrast, errors in $\kappa$ from the designed value reduce the efficiency of conventional adiabatic and directional couplers, and these errors \textit{cannot be corrected after fabrication}.
\end{itemize}

\textbf{\textit{Simulation}}: We conducted 3D finite-difference time-domain simulations (Ansys Lumerical FDTD) of an example physical implementation of the SAPCB shown in Figure 3 and using the  parameters in Table \ref{sim_params}. These simulations assumed high-index single-mode polymer core  SAPCB waveguides embedded in a low-index fluoropolymer cladding, and silicon nitride (SiN) PIC waveguides in silicon dioxide cladding. SiN is a high-index contrast waveguide platform transparent over the visible and infrared and is available in most silicon photonics and CMOS foundries \cite{blumenthal_silicon_2018}. The simulations assumed a wavelength $\lambda = 1550$ nm, and the optimized design exhibits less than 0.2 dB insertion loss.

\begin{table}
\caption{Simulation parameters }
\label{sim_params}
\begin{tabular}{  m{4cm}| m{3cm}  } 
    Polymer core $n$ \cite{rabiei_polymer_2002} & 1.575 \\
    \hline
    Polymer cladding $n$ \cite{lacraz_femtosecond_2015} & 1.34 \\
    \hline
    Polymer core $\mathrm{d}n/\mathrm{d}T$ \cite{rabiei_polymer_2002} & $-1.1 \times 10^{-4}~/^\circ \mathrm{C}$\\\hline Polymer cladding $\mathrm{d}n/\mathrm{d}T$ \cite{lacraz_femtosecond_2015} & $-5 \times 10^{-5}~/^\circ \mathrm{C}$\\\hline
    PIC waveguide core $n$ \cite{elshaari_thermo-optic_2016} & 2 \\
    \hline
    PIC waveguide cladding $n$ \cite{elshaari_thermo-optic_2016} & 1.445 \\
    \hline
    PIC core $\mathrm{d}n/\mathrm{d}T$ \cite{elshaari_thermo-optic_2016} & $2.51 \times 10^{-5}~/^\circ \mathrm{C}$\\\hline PIC cladding $\mathrm{d}n/\mathrm{d}T$ \cite{elshaari_thermo-optic_2016} & $9.6 \times 10^{-6}~/^\circ \mathrm{C}$\\\hline
    SiN width & 462.5 nm \\
    \hline
    Polymer width & 1.6 $\mu$m \\
    \hline
    SiN height & 300 nm \\
    \hline
    Polymer height & 1 $\mu$m \\
    \hline
    Gap ($g$) & 1 $\mu$m \\
    \hline
    Length ($L$) & 100 $\mu$m \\
    \hline
    $\theta_\text{opt}$ & $4.4^{\circ}$\\\hline
    Wavelength ($\lambda$) & 1550 nm
\end{tabular}
\end{table}

\begin{figure*}[t]
    \centering
    \includegraphics[width=\textwidth]{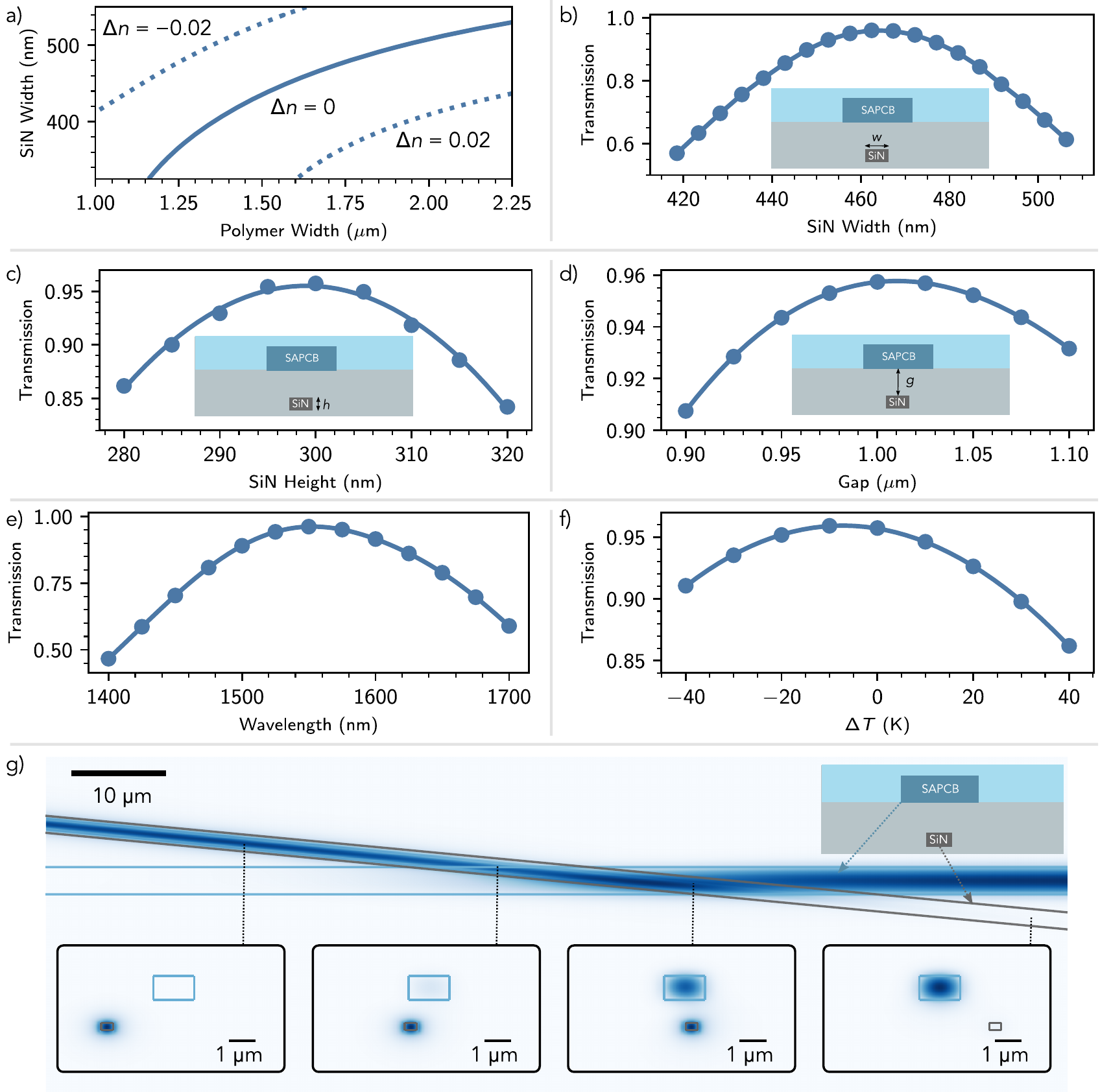}
    \caption{a) Effective mode index mismatch $\Delta n = n_\text{SAPCB} - n_\text{PIC}$ as a function  of the PIC (SiN) waveguide width and the SAPCB (polymer) waveguide width. The two waveguide geometries should be engineered such that their modes have equal propagation constants, i.e. $\Delta k = 2\pi \Delta n / \lambda = 0$. b-f) Power transfer efficiency as a function of the PIC waveguide width (b), PIC waveguide height (c), coupling gap (d), wavelength (e), and temperature (f) for the design with parameters in Table I. g) The field profile of the alignment-free coupler with parameters in Table I. The insets below show the cross-sectional field profile at varying points along the propagation.}
    \label{sim_results}
\end{figure*}

Figure \ref{sim_results}a plots the effective mode index mismatch $\Delta n = n_\text{SAPCB} - n_\text{PIC}$ as a function of the SiN and polymer waveguide widths. Efficient mode transfer between the waveguides requires matching their propagation constants by engineering their geometry. This  requirement dominates the fabrication-induced error. Figures \ref{sim_results}b-f plot the effect on transmission caused by errors in SiN width (b), SiN height (c), coupling gap (d), wavelength (e), and temperature (f). The coupler is remarkably robust to changes in all of these parameters, exhibiting less than 0.5 dB penalty for a $\pm$ 20 nm variation in waveguide dimensions and lower than 0.3 dB excess loss for a $\pm$ 100 nm change in the coupling gap $g$. Moreover, it has a 1-dB optical bandwidth in excess of 180 nm and exhibits less than 0.5 dB temperature sensitivity over a range of 80$^\circ$ C. We obtained the results in Figure 3 from FDTD simulations of the full structure. Fig.~\ref{sim_results}g shows these simulations for the parameters of Table \ref{sim_params}, illustrating the power transfer from the SiN waveguide,  through the alignment-free coupler, and into the polymer waveguide. Cross sectional field intensity plots along the structure are shown underneath. 

\begin{figure}
    \centering
    \includegraphics[width=\columnwidth]{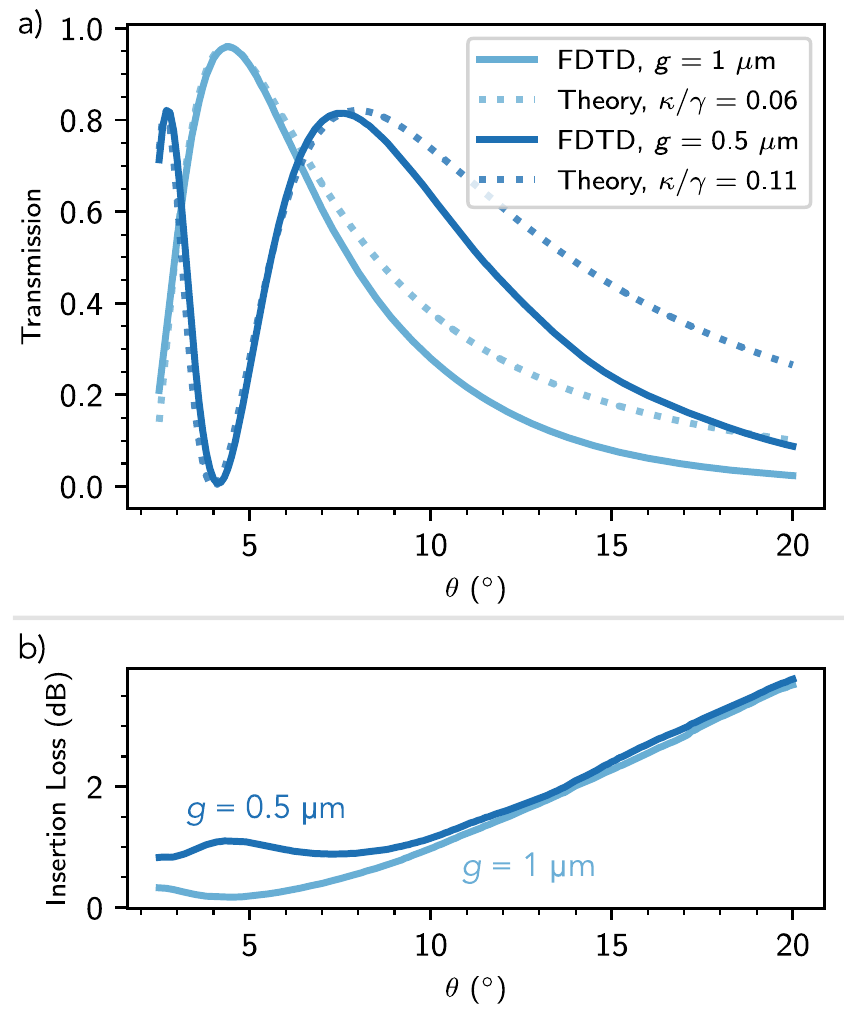}
    \caption{a) Power transfer efficiency $\eta$ vs $\theta$ for designs with coupling gap $g = 1~\mu$m and $g = 0.5~\mu$m. The solid lines indicate FDTD simulation results, while the dotted lines are fit to equation (\ref{theory}). b) Insertion loss $(|a_\text{out}|^2 + |A_\text{out}|^2)/|a_\text{in}|^2$ as a function of $\theta$. At large angles, there is significant radiation loss due to the non-adiabatic nature of the intersection.}
    \label{angle_plot}
\end{figure}

Figure \ref{angle_plot}a plots $\eta$ vs. $\theta$ for the same structure. The waveguide intersection causes a scattering loss of $\sim$0.2 dB at the optimal coupling angle $\theta$.  Upon correcting for this loss, $\eta$ agrees well with the expression in Eq.~\ref{theory} around this region and exhibits an angular alignment (3 dB) tolerance $\Delta \theta > 5$ degrees. Additionally, $\eta$ rolls off slowly for $\theta > \theta_\text{opt}$, permitting modest coupling efficiencies at even large angular errors. This feature of our approach greatly simplifies initial alignment and relaxes the required precision of alignment during packaging. 

The scattering loss shown in Figure 4a results primarily from a  faster-than-adiabatic transition at the waveguide intersection and  increases with $\theta$ as the transition into the hybridized modes becomes more abrupt \cite{xia_mode_2006, spillane_ideality_2003}. This loss is particularly significant at large $\theta$, shown in Figure \ref{angle_plot}b, which accounts for the discrepancy compared to theory (eq. \ref{theory}) at these angles. The scattering loss drops with increasing $g$, which makes the transition more adiabatic. Increasing $g$ introduces two tradeoffs, however: the angular tolerance $\Delta \theta$ will drop, and the transmission will be more sensitive to errors in $\Delta k$. If higher insertion losses are acceptable, the waveguides can be coupled more strongly, which improves $\Delta \theta$. We also show in Figure \ref{angle_plot}a such an example, where we decreased $g$ to 500 nm.  $\kappa/\gamma$, and therefore the tolerance, $\Delta \theta$, nearly doubles, but at the expense of a higher insertion loss of 1 dB. The tradeoffs between insertion loss, robustness to fabrication error, and $\Delta \theta$ bound an optimal range for $\kappa$ and therefore $g$.

\begin{figure}
    \includegraphics[width=\columnwidth]{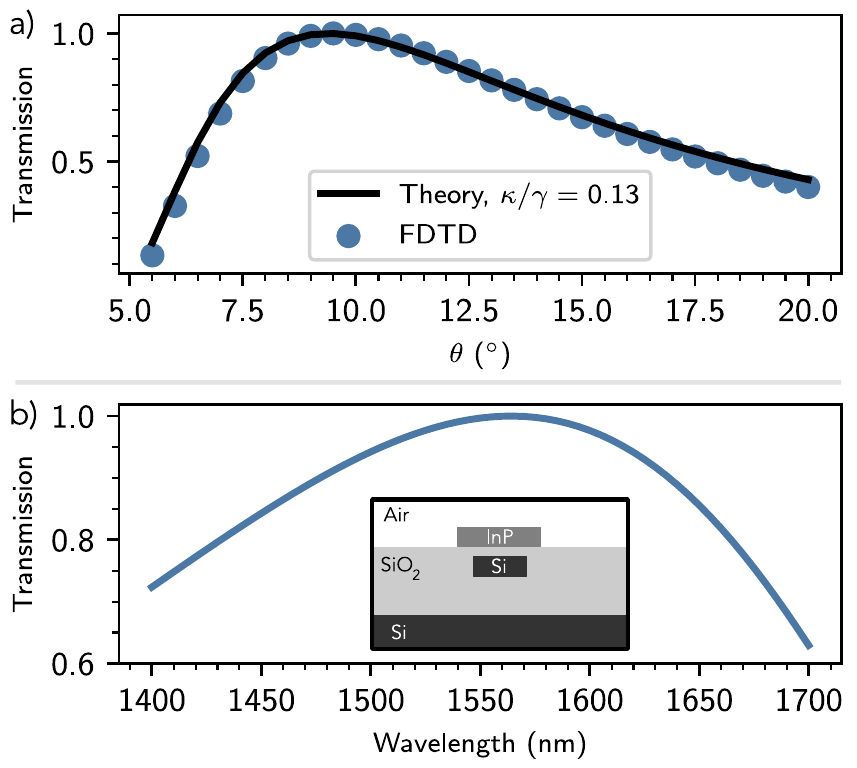}
    \caption{a) Transmission vs. $\theta$ for an alignment-free coupler designed to interface a 640 $\times$ 300 nm InP gain microchiplet to a 500 $\times$ 220 nm silicon photonic waveguide. The strong mode confinement in both materials eliminates scattering loss at the intersection, permitting mode transfer with no insertion loss. As a result, the transmission characteristic reproduces nearly perfectly equation (\ref{theory}). b) The transmission efficiency as a function of wavelength. The coupler has a 1-dB bandwidth exceeding 230 nm.}
    \label{chiplet_plot}
\end{figure}

Insertion losses can also be reduced by employing higher-index platforms that enable stronger vertical confinement of the optical mode. Figure \ref{chiplet_plot} shows such an example, where we design a coupler to interface a 500 $\times$ 220 nm  silicon photonic waveguide to a 640 $\times$ 300 nm indium phosphide (InP) waveguide ($g = 150$ nm) for hybrid integration of gain. Silicon and InP have much higher refractive indices ($n_\text{Si} = 3.47$; $n_\text{InP} = 3.17$) and therefore confine the optical mode more strongly; as a result, the optical mode is significantly less perturbed by the introduction of the other waveguide at the intersection. As Figure \ref{chiplet_plot}a shows, this enables efficient mode transfer \textit{with no insertion loss}; moreover, we find a near-exact fit to theory. Additionally, despite the high index contrast of both the Si and InP photonic platforms, which would imply they are strongly dispersive, we find that our optimized coupler has a 1-dB optical bandwidth exceeding 230 nm (Fig.~\ref{chiplet_plot}b).

\section*{Discussion}

In Figure 6, we compare the lateral and angular alignment tolerance of the alignment-free coupler to a 10 $\mu$m inverse tapered edge coupler and a tapered adiabatic coupler. For each approach, we compare the 1-dB coupling efficiency contour in the $\delta r_{||}$-$\delta \theta$ plane to that of the alignment-free coupler. 

Non-perturbative approaches, such as edge coupling (EC), have a fundamental tradeoff between the lateral and angular tolerances in coupling efficiency. Assuming the mode $E_\text{wg}(\vec{r})$ produced when the waveguide couples into free space is Gaussian, one can calculate the mode overlap \cite{papes_fiber-chip_2016}
\begin{equation}
    \eta_\text{EC} = \frac{\lvert \int E_\text{wg}^*(\vec{r}) E_\text{fiber}(\vec{r})~\mathrm{d}A \rvert ^2}{\int \lvert E_\text{wg}(\vec{r})\rvert^2 ~\mathrm{d}A \int \lvert E_\text{fiber}(\vec{r})\rvert^2 ~\mathrm{d}A}
\end{equation}
with the input fiber mode $E_\text{fiber}(\vec{r})$, also assumed to be Gaussian but misaligned by an angle $\delta \theta$ and transverse distance $\delta r_{||}$. We assume that $E_\text{fiber}(\vec{r})$ and $E_\text{wg}(\vec{r})$ have identical beam waist radius $w_0$ to maximize mode-matching; therefore, $\eta_\text{EC}$ is unity when there is no misalignment. Assuming a small angular error $\delta \theta$ in the paraxial limit, we find that:
\begin{equation}
    \eta_\text{EC}(\delta r_{||}, \delta \theta) = \exp \left (-\frac{|\delta r_{||}|^2}{w_0^2} - \frac{k^2 w_0^2}{4}  \delta \theta^2 \right )
\end{equation}
This translates to a fundamental tradeoff between the lateral ($\Delta r_{||}$) and angular ($\Delta \theta$) alignment tolerances:
\begin{equation}
    \Delta r_{||} \Delta \theta =  w_0 \left ( \frac{2}{k w_0} \right ) = \frac{2}{k} = \frac{\lambda}{\pi n}
\end{equation}

This tradeoff does not apply to the alignment-free coupler, which  has both a high angular tolerance and an arbitrarily high lateral tolerance that can be increased by increasing $L$. As a result, the combined lateral and angular tolerance $\Delta r_{||} \Delta \theta$ of our approach exceeds the fundamental limit on alignment tolerances for edge coupling. Expanding or contracting the beam size improves the alignment tolerance of edge coupling in one dimension at the expense of the other; thus, no possible edge coupler can have \textit{both} superior lateral and superior angular tolerance to that of an alignment-free coupler.

Adiabatic couplers, on the other hand, taper one or both waveguides to induce an avoided crossing between the two eigenmodes, which adiabatically transfers power from one waveguide to the other \cite{sun_adiabaticity_2009, ramadan_adiabatic_1998}. This adiabatic transition makes the devices robust to variation in $\Delta k$, which has led to them being favored in many photonic platforms for their resilience to fabrication error. This robustness comes at the cost of alignment tolerance, however, as small lateral or angular errors render the interaction non-adiabatic, resulting in little or no power transfer. To compare relative tolerances, we designed an adiabatic coupler to transfer power from SiN to the polymer waveguide; our coupler linearly tapers the SiN waveguide width from 550 to 320 nm over a length of 200 $\mu$m ($g = 1~\mu$m) and achieves an efficiency of 96\%, which is comparable to our optimized alignment-free coupler. 

Figure 6 shows the transmission penalty of the adiabatic coupler as a function of misalignment $\delta r_{||}, \delta \theta$; while $\Delta \theta$ is comparable to an alignment-free coupler of the same length, $\Delta r_{||}$ is far smaller. For sufficiently long tapers, it is also possible to achieve high coupling efficiency at a large angular error $\delta \theta$, where the taper acts effectively as an alignment-free coupler. We omit this region in Figure 6, as the coupling in this regime is non-adiabatic. Moreover, as the adiabatic coupler is tapered, unlike the alignment-free coupler, the lateral tolerance $\Delta r_{||}$ in this regime is far smaller.

\begin{figure}
    \includegraphics[width=\columnwidth]{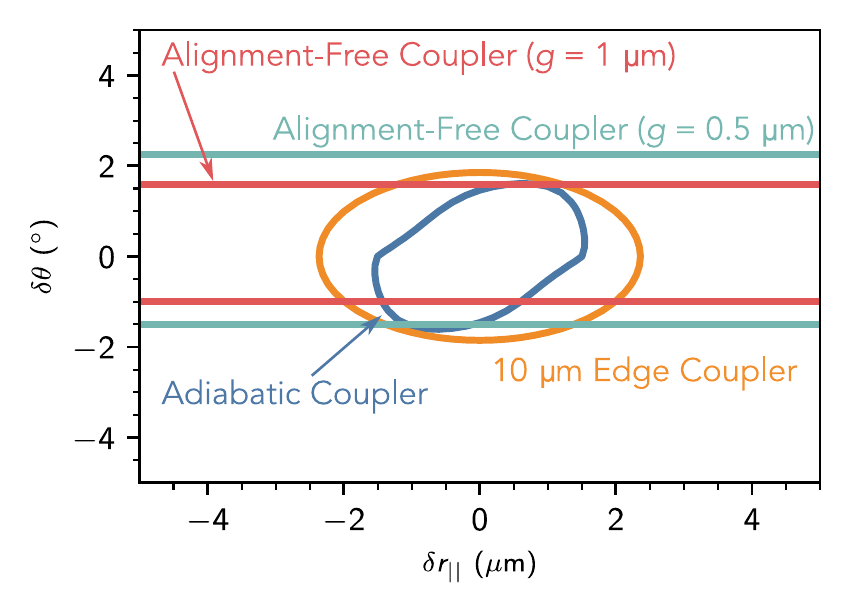}
    \caption{Lateral and angular alignment tolerance of the alignment-free coupler compared to inverse tapered edge couplers and tapered adiabatic couplers. The lines indicate the 1-dB coupling efficiency contour as a function of in-plane displacement $\delta r_{||}$ and angular displacement $\delta \theta$. The alignment-free coupler has a combined alignment tolerance $\Delta r_{||} \Delta \theta$ that exceeds current approaches.}
\end{figure}

Our results show that the combined lateral and angular tolerance $\Delta r_{||} \Delta \theta$ of our approach is  higher than that of conventional optical couplers. We achieve this by making use of evanescent coupling, which does not suffer from a fundamental limitation on $\Delta r_{||} \Delta \theta$, and by intentionally engineering a system largely invariant to lateral displacements. This lateral tolerance is maximized by choosing both waveguides to not be tapered; as a result, $\Delta r_{||}$ can be arbitrarily high. While our approach does require the effective indices of the waveguides to be matched,  our simulation results suggest that the dimensional tolerances needed to achieve this are well below what is realizable in current fabrication processes.

\begin{figure}
    \includegraphics[width=\columnwidth]{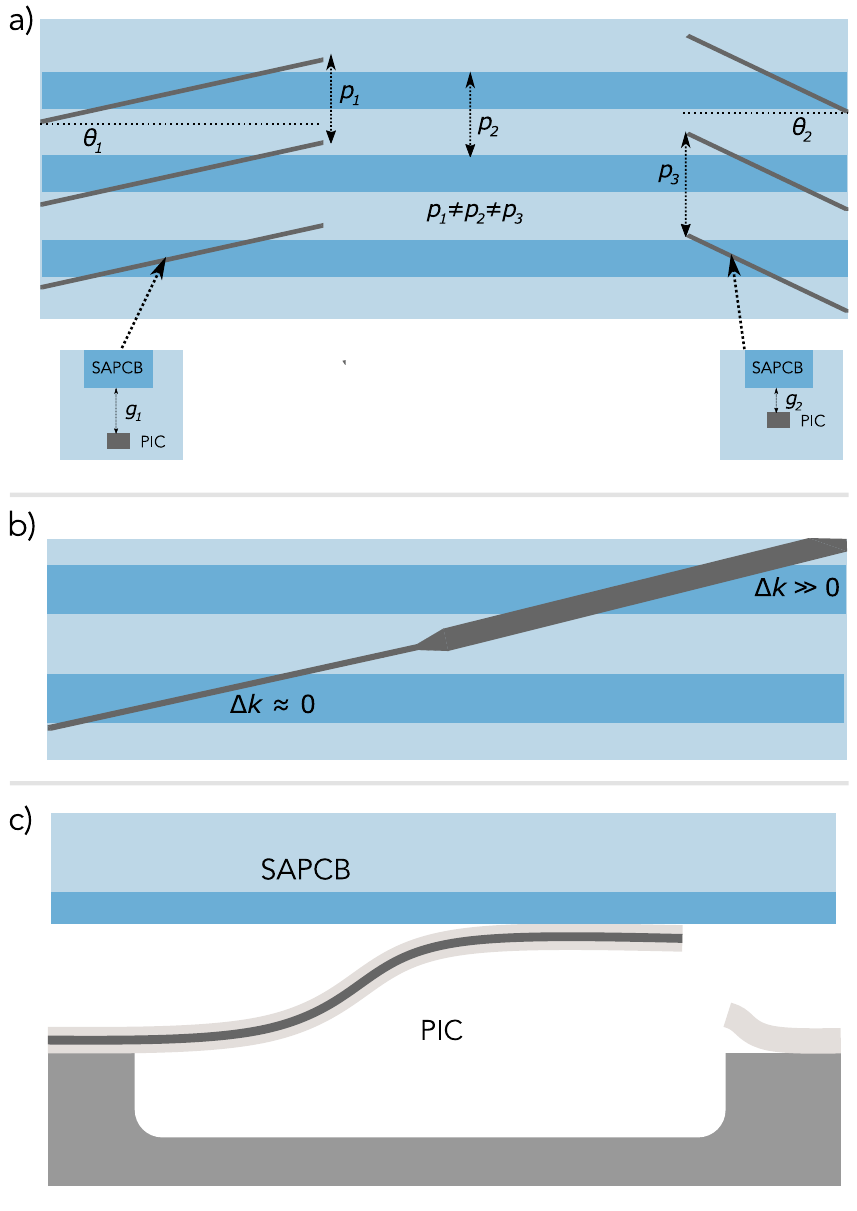}
    \caption{a) The alignment-free coupler can interface photonic circuits with differing waveguide pitches and process stacks. As the waveguides interact at an angle, precise matching of the waveguide pitch is not necessary. By varying the coupling angle $\theta$, one can easily optimize the transmission for any coupling gap $g$. b) The requirement for phase-matching permits simplified routing with minimal crosstalk. By tapering the waveguide to ensure $\Delta k \gg 0$, waveguides can be routed over one another with negligible crosstalk. c) Cantilevering one or both waveguides enables automatic vertical alignment in addition to in-plane tolerance.}
    \label{applications}
\end{figure}

\textbf{\textit{System Integration and Outlook}}: We show in Figure \ref{applications} two possible applications of the SAPCB for system-level integration. In Figure \ref{applications}a, we consider interfacing two photonic circuits with different waveguide pitches and process stacks. As the alignment-free coupler interfaces with the polymer waveguides at an angle, the off-axis intersection guarantees that both waveguides can couple into the same waveguide on the board. The two PICs may also not have the same process stack; for instance, one process may require a larger oxide layer, resulting in a larger coupling gap $g$ to the polymer waveguide. This can be addressed in our approach by simply modifying the coupling angle $\theta$ to preserve efficient power transfer. 

Figure \ref{applications}b demonstrates another advantage originating from the need for $\Delta k \approx 0$ for efficient coupling. Suppose a waveguide on a PIC needs to be routed over a polymer waveguide with minimal crosstalk. By engineering the dimensions of the PIC waveguide, one can ensure a strong wavevector mismatch $\Delta k$ with the polymer waveguide, allowing for crosstalk-free transmission of signals over many photonic components on a board.

Alignment tolerance can also be extended to the third (vertical) dimension. Our approach couples chips and substrates to one another vertically, which is efficient when surfaces are planar; however, flatness is not always guaranteed. For example, two chips may not be polished well enough, they may have necessary protrusions, or there may be contamination between the chips that prevents a contact reliable enough for evanescent mode transfer. For these situations, cantilevering one or both couplers can enable alignment tolerance in the vertical dimension. 

As illustrated in Fig.~\ref{applications}c, the cantilever may consist of an oxide-clad waveguide layer, which is released by defining a trench around it and using an isotropic chemical etch for a selective release step \cite{sun_cantilever_2009, wood_compact_2012}. If the waveguide layer is strained in tension, it will curl the oxide membrane up. Upon stacking the two chips, the cantilever comes to rest spring-loaded between the chips, providing automatic vertical alignment in addition to the existing in-plane alignment. Moreover, we are left with an alignment platform that is resilient to modest translations in all three dimensions: $x$, $y$, and $z$. The cantilever can be singly clamped, as illustrated in the figure, or doubly clamped for stiffer, but smaller, out of plane displacement. Previous demonstrations of cantilevered couplers have interfaced directly to optical fiber \cite{sun_cantilever_2009, yoshida_vertically_2016, wood_compact_2012}, posing a similar alignment challenge to conventional edge couplers; however, the alignment-free coupler greatly simplifies interfacing to these structures.

We envision that the alignment-free coupler would be defined on the PIC, where advanced photolithography processes  define the required waveguide geometry and angle precisely. This frees the SAPCB to consist only of linear arrays of polymer waveguides, with no bends or tapering required. The simple layout of the polymer board could potentially allow it to be fabricated by fiber pulling approaches from a preform, rather than more costly lithography processes. Additionally, polymer waveguides have a wide transparency window, making the SAPCB applicable to photonics operating in both the visible and near-infrared. 

\textbf{\textit{Conclusion}}: We have presented a self-aligning photonic circuit board capable of serving as a universal connector for optoelectronic system integration. The critical element of the SAPCB is the alignment-free coupler, which engineers a laterally invariant system insensitive to the exact location of waveguides on the photonics, and which also exhibits high angular tolerance and arbitrarily high lateral tolerance. Our approach is robust to variations in the device geometry, and we show its combined lateral and angular tolerance exceeds that of conventional optical coupling approaches. The SAPCB allows for system integration with minimal design and alignment requirements, enabling a wide range of photonic components to interface with one another and simplifying the assembly of complex optical systems.\\
\\
\textbf{Competing Interests:} The authors have filed US Patent Application 17/470,803.\\
\\
\textbf{Data Availability:} The data that support the plots in this paper and other findings of this study are available from the authors upon reasonable request.\\
\\
\textbf{Author Contributions:} D.E. conceived the project. S.B. conducted FDTD simulations of the structure. S.B. and D.E. derived the analytical models and wrote the manuscript.\\
\\
\textbf{Acknowledgments:} S.B. was supported by a National Science Foundation (NSF) Graduate Research Fellowship under grant no. 1745302 and the Air Force Office of Scientific Research (AFOSR) under award numbers FA9550-16-1-0391 and FA9550-20-1-0113. D.E. acknowledges support from the Air Force Research Laboratory (AFRL) under award number FA8750-20-2-1007. The authors are grateful to Dr. Carlos Errando Herranz, Dr. Mohamed ElKabbash, Dr. Genevieve Clark, and Alexander Sludds for useful discussions.

\bibliographystyle{osajnl}
\bibliography{Bibliography}

\end{document}